\documentclass[twocolumn,showpacs,prc,aps,floatfix,hidelinks]{revtex4-2}

\usepackage{bm}
\usepackage{braket}
\usepackage{amssymb}
\usepackage{graphicx}
\usepackage{dcolumn}
\usepackage{amsmath}
\usepackage{amsfonts}
\usepackage{xcolor}
\usepackage{hyperref}
\usepackage{multirow}
\usepackage{tabularx}

\usepackage{mathtools}
\mathtoolsset{showonlyrefs}

\begin{document}

\title{\textit{Ab initio} calculation of the $\beta$-decay from $^{11}$Be to a p${+}^{10}$Be resonance}
\author{M. C. Atkinson$^1$, P. Navr\'atil$^1$, G.~Hupin$^2$, K.~Kravvaris$^3$, S.~Quaglioni$^3$} 
\affiliation{${}^1$TRIUMF, Vancouver, British Columbia, V6T 2A3, Canada}
\affiliation{${}^2$Université Paris-Saclay, CNRS/IN2P3, IJCLab, 91405 Orsay, France}
\affiliation{${}^3$Lawrence Livermore National Laboratory, P.O. Box 808, L-414, Livermore, CA 94551, USA}

\date{\today}

\begin{abstract}
The exotic $\beta$-delayed proton emission is calculated in $^{11}$Be from first principles using chiral two- and three-nucleon forces. To investigate the unexpectedly-large branching ratio measured in [PRL 123, 082501 (2019)] we calculate the proposed $(1/2^+,1/2)$
proton resonance in $^{11}$B using the no-core shell model with continuum. This calculation helps to address whether this enhancement is caused by unknown dark decay modes or an unobserved proton resonance. We report a branching ratio of $b_p = (1.3\pm0.5)\times10^{-6}$, suggesting that its unexpectedly-large value is caused by an unobserved proton resonance in $^{11}$B.
\end{abstract}

\maketitle

\section{Introduction}

Nuclear $\beta$-decay is well-recognized as a process sensitive to nuclear
structure~\cite{Bohr-Mottelson}
. When considering neutron-rich nuclei, a particularly interesting
$\beta$-decay reaction known as $\beta$-delayed particle
emission is possible~\cite{Baye:2011}. Specifically, $\beta$-delayed proton emission is a rare process in
which the parent nucleus undergoes $\beta$-decay into a proton-unbound state from
which the proton is emitted.
Due to energy conservation, this exotic process is forbidden unless $S_n < (m_n-m_p-m_e)c^2 \approx 782$ keV where $S_n$ is the neutron separation energy and $m_n$, $m_p$, and $m_e$ are the neutron, proton, and electron rest masses, respectively~\cite{Baye:2011}.
The $^{11}$Be nucleus is a halo nucleus with a
neutron separation energy of 0.5016 MeV, making it an ideal candidate for such a decay. It has been
suggested that this particular process can shed some light on the neutron lifetime
puzzle in that it could reveal an exotic dark-decay mode~\cite{Volya:2020}.

The $\beta$ decay of the ground state of $^{11}$Be into the continuum was originally calculated using a cluster expansion resulting in a
small branching ratio, $b_p = 3.0\times 10^{-8}$~\cite{Baye:2011}. This small branching suggested it was extremely rare compared to other $\beta$-decay branches such as to the ground state ($b_p = 0.547$) or to one of the many excited states in $^{11}$B (e.g. $b_p = 0.314$ to the $1/2_1^-$ excited state, $b_p = 0.0647$ to the $1/2_1^+$ excited state etc.)~\cite{Millener:1982}.
To further investigate this $\beta$-delayed proton emission, Riisager \textit{et
al.} indirectly measured the decay of $^{11}$Be to $^{10}$Be~\cite{Riisager:2014}.  
This experiment revealed a large branching ratio of $b_p = 8.3(9)\times 10^{-6}$,
two orders of magnitude larger than the previous theoretical calculation.  Thus, two
possible explanations were proposed for this discrepancy. Either the halo neutron
decays to an unobserved proton resonance in $^{11}$B, or there are other unobserved,
exotic neutron decay modes such as dark decay modes~\cite{Riisager:2014}. 
Seven years later, in
2019, Ayyad \textit{et al.} directly observed the $^{11}$Be $\beta$-delayed proton
emission in an experiment performed at TRIUMF. The corresponding experimental branching
ratio of $b_p=1.3(3)\times 10^{-5}$ is consistent with the large branching ratio
observed in the previous indirect experiment~\cite{Ayyad:2019}. This branching ratio equates to
$B(\mathrm{GT}) = 5.5^{+8.3}_{-3.3}$ which falls under the theoretical limit of 3 for the $\beta$-decay of a free neutron within one standard deviation~\cite{AyyadErratum:2020}. Furthermore, because this
experiment involved direct detection, the ejected proton distribution was used to locate the possible proton resonance in $^{11}$B. The resonance was found to have spin 1/2 (or 3/2), positive parity, and isospin 1/2 with an excitation energy of
197 keV.  This measurement supported the first explanation: the existence of an
unobserved resonance in $^{11}$B.

There have been several theoretical investigations of the proposed $^{10}$Be+$p$ resonance with differing results. 
The authors of Ref.~\cite{marek:2021}, using a shell model embedded in the continuum (SMEC), conclude that the branching ratio of $\beta$-delayed $\alpha$ emission, the predicted resonance width, and the observed $\beta$-delayed proton emission in Ref.~\cite{Ayyad:2019} can not be reconciled within their calculations.
Another shell model analysis of this system claims that the experimentally observed branching ratio is impossible to explain~\cite{Volya:2020}. The halo effective field theory analysis in Ref.~\cite{Elkamhawy:2021} supports the existence of the proposed proton resonance.

In this work, we address the $\beta$-delayed proton emission in $^{11}$Be from first principles using the 
no-core shell model with continuum (NCSMC)~\cite{Baroni2013L,Simone:2013,PhysRevC.90.061601,Navratil:2016}. First, we carry out $p$+$^{10}$Be scattering calculations to search for the proposed resonance in $^{11}$B. Second, we develop the framework to consistently compute $\beta$-decay matrix elements within the NCSMC approach and evaluate the $\beta$-decay branching ratio from the ground state of $^{11}$Be to the $p$+$^{10}$Be channel.
The NCSMC is well-suited to describe this particular decay since it not only accurately describes the structure of light nuclei, but properly profiles the continuum in the low-energy
regime.  In Sect.~\ref{theory}, we briefly introduce the NCSMC and the microscopic Hamiltonian adopted in the present study. Main results are presented in Sect.~\ref{results} and conclusions are given in Sect.~\ref{concl}. Details of the calculation of the NCSMC formalism for the calculation of $\beta$-decay matrix elements are given in Appendix~\ref{sec:appendix}.

\section{Theory}\label{theory}

The $\beta$-decay operator of the $J^\pi{=}1/2^+$, $T{=}3/2$ ground state of $^{11}$Be to a $J^\pi{=}1/2^+$,$T{=}1/2$ proton-unbound $p+^{10}$Be state
is purely driven by the (reduced) matrix elements of the Gamow-Teller (GT) operator due to the change in isospin~\cite{Bohr-Mottelson}:
 \begin{equation}
    B(\mathrm{GT}) =
    \frac{1}{2}\left|\Braket{\Psi_{^{11}B}^{\frac{1}{2}^+\frac{1}{2}}||\hat{\mathrm{GT}}||\Psi_{^{11}Be}^{\frac{1}{2}^+\frac{3}{2}}}\right|^2.
    \label{eq:bgt}
 \end{equation}
Here we consider the leading order GT operator
\begin{equation}
\hat{\mathrm{GT}} = \sum_{i=1}^A \hat{\bm{\sigma}}_i\hat{\tau}^+_i,
 \label{eq:sigma-tau}
\end{equation}
where $\hat{\bm{\sigma}}_i$ is the single-particle Pauli operator and $\hat{\tau}_i^+$ is the single-particle isospin-raising operator~\cite{Bohr-Mottelson}.
The evaluation of Eq.~\eqref{eq:bgt} requires a framework such as the NCSMC where bound and scattering wave functions are consistently calculated.
%
%
%
The ansatz for the NCSMC initial(final) state is a generalized
cluster expansion~\cite{Navratil:2016}

\begin{align}
   \Ket{\Psi^{J^\pi T}_A} = &\sum_\lambda c_\lambda^{J^\pi T}\Ket{A \lambda J^\pi T} \nonumber\\
   &+ \sum_\nu\int drr^2\frac{\gamma_{\nu}^{J^\pi T}(r)}{r}\hat{\mathcal{A}}_\nu\Ket{\Phi_{\nu r}^{J^\pi T}}.
   \label{eq:ncsmc}
\end{align}
The first term is an expansion over no-core shell model (NCSM)~\cite{Barrett:2013} eigenstates of the aggregate system
$\Ket{AJ^\pi T}$ (either $^{11}$Be in the intial state or $^{11}$B in the final state) calculated in a many-body harmonic oscillator basis.
The second term is an expansion over microscopic cluster channels
$\hat{\mathcal{A}}_\nu\Ket{\Phi_{\nu r}^{J^\pi T}}$ which describe the clusters (either
$^{10}$Be+$n$ in the initial state or $^{10}$Be+$p$ in the final state) in relative motion:
\begin{align}
    \ket{\Phi^{J^\pi T}_{\nu r}} = &\Big[ \big( \ket{^{10} {\rm Be} \, \alpha_1 I_1^{\pi_1} T_1} \ket{N \, \tfrac12^{\texttt{+}}\tfrac12} \big)^{(sT)}
     Y_\ell(\hat{r}_{10,1}) \Big]^{(J^{\pi}T)} \nonumber \\
    &\times\,\frac{\delta(r{-}r_{10,1})}{rr_{10,1}} \; ,
    \label{eq:Be10N_rgm_state}
\end{align}
where 
$\ket{^{10} {\rm Be} \, \alpha_1 I_1^{\pi_1} T_1}$ and 
$\ket{N \, \tfrac12^{\texttt{+}}\tfrac12}$ are the eigenstates of $^{10}$Be and $N$, respectively, 
with $N$ representing $n$ for the initial state ($^{11}$Be) and $p$ for the final state ($^{11}$B) . 
The cluster channels enable the description of scattering states as well as weakly bound extended (halo) states in the NCSMC.
The $^{10}$Be eigenstate (also calculated within the NCSM) has angular momentum $I_1$, parity $\pi_1$, isospin $T_1$, and energy label $\alpha_1$.
Here $r$ denotes the distance between the clusters and $\nu$ is a collective index of
the relevant quantum numbers. The coefficients $c_\lambda^{J^\pi T}$ and
relative-motion amplitudes $\gamma_{\nu}^{J^\pi T}(r)$ are found by solving a
two-component, generalized Bloch Schr\"{o}dinger equations derived in detail
in Ref.~\cite{Navratil:2016}.
The $\hat{\mathcal{A}}_\nu$ term is the inter-cluster antisymmetrizer:
\begin{align}
\hat{\mathcal{A}}_\nu = \sqrt{\frac{(A-1)!}{A!}}\left(1+\sum_{P\neq id}(-1)^pP\right),
\label{eq:antisymmetrizer}
\end{align}
where the sum runs over all possible permutations of nucleons P (different from the identical one) that can be carried out between the target cluster and projectile, and $p$ is the number of interchanges characterizing them.
The resulting NCSMC equations are solved using the
coupled-channel R-matrix method on a Lagrange mesh~\cite{Baye:2010,Simone:2013}.

We start from a microscopic Hamiltonian including the nucleon-nucleon ($NN$) chiral interaction at
next-to-next-to-next-to-next-to leading order (N$^{4}$LO) with a cutoff
$\Lambda{=}500$ MeV developed by Entem \textit{et al}~\cite{Entem:2015,Entem:2017}, denoted as
$NN$-N$^{4}$LO(500). In addition to the two-body interaction, we include a
three-body ($3N$) interaction at next-to-next-to leading order (N$^{2}$LO) with
simultaneous local and nonlocal regularization ~\cite{Navratil:2007,Gennari:2018,Gysbers2019NatPhys,Soma2020}.
The whole interaction (two- and three-body) will be referred to as $NN$-N$^{4}$LO(500)$+3N_{\rm lnl}$.
A faster convergence of our NCSMC calculations is obtained by softening the Hamiltonian through
the similarity renormalization group (SRG)
technique~\cite{Wegner1994,Bogner2007,PhysRevC.77.064003,Jurgenson2009}.
The SRG unitary transformation induces many-body forces that we include up to the three-body level.
Four- and higher-body induced terms are small at the $\lambda_{\mathrm{SRG}}{=}1.8$ fm$^{-1}$
resolution scale used in the present calculations~\cite{PhysRevC.103.035801}. Concerning the frequency of the underlying HO basis, we choose $\hbar\Omega = 18$ MeV for which the ground state energies of the investigated nuclei present minimum. 
For technical reasons, we are not able to reach basis sizes beyond $N_{\rm max}{=}7$ for $^{11}$B and $^{11}$Be.

 We note that the present calculations are the first application of the NCSMC approach to the description of $\beta$-decay transitions. While the present calculation is implemented solely for the GT operator, the formalism is also valid for the Fermi (as well as the spin part of M1) one-body operator. We are able to calculate the relevant matrix elements without approximations (as opposed to the radiative capture calculations in, e.g., Ref.~\cite{PhysRevC.103.035801}) and evaluate the transition kernels (i.e., matrix elements entering the integrals in Eq.~\eqref{eq:four-terms}) using a similar technique applied to calculate Hamiltonian interaction/norm kernels. See Appendix~\ref{sec:appendix} for the full derivation of the NCSMC GT matrix element utilizing second quantization.
 
\section{Results}\label{results}
\subsection{NCSMC calculations for $^{11}$Be and $^{11}$B}

\begin{figure}[h]
   \begin{center}
      \includegraphics[scale=1.2]{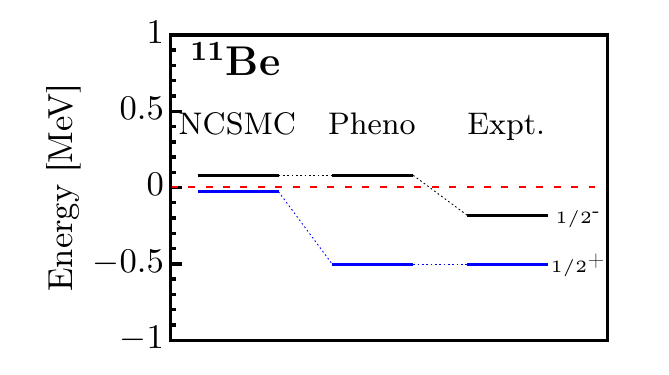}
   \end{center}
   \caption{Calculated and experimental levels of $^{11}$Be. Only states corresponding to experimentally bound states with respect to the $^{10}$Be$+$n threshold (horizontal red dashed line) are shown. The left column shows the original {\it ab initio} NCSMC calculation in $N_{\rm max}{=}7$ space. The phenomenological adjusted calculation is presented in the middle column. See the text for details.}
   \label{fig:levels_be11}
 \end{figure}

\begin{figure}[h]
   \begin{center}
      \includegraphics[scale=1.2]{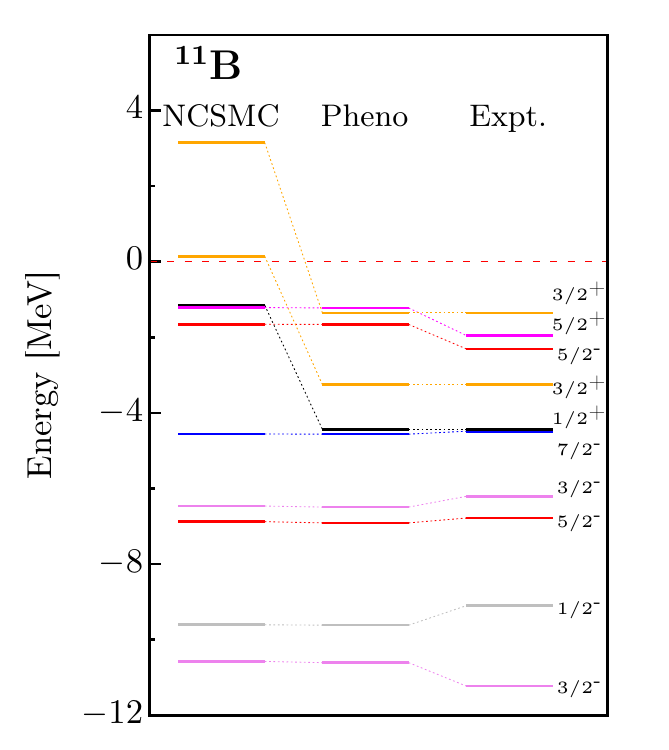}
   \end{center}
   \caption{Calculated and experimental levels in $^{11}$B. Selected states corresponding to experimentally bound states with respect to the $^{10}$Be$+$p threshold (horizontal red dashed line) are shown. The SRG evolved $NN$-N$^{4}$LO(500)$+3N_{\rm lnl}$ interaction was used. The left column shows the original {\it ab initio} NCSMC calculation in $N_{\rm max}{=}7$ space. 
   The phenomenolgically adjusted calculation is presented in the middle column. 
   See the text for details.
   The isospin of all shown states is $T{=}1/2$.} 
   \label{fig:levels_b11}
\end{figure}

We start by performing NCSM calculations for $^{10,11}$Be and $^{11}$B. The obtained eigenvalues and eigenvectors serve as input for the NCSMC. For the expansion in Eq.~\eqref{eq:ncsmc}, we used the $^{10}$Be $0^+$ ground state and the first excited $2^+$ state, the lowest 12 (10) positive (negative) parity eigenstates for $^{11}$Be, and 20 (12) positive (negative) eigenstates for $^{11}$B with $
J$ ranging from $1/2$ to $11/2$. The resulting NCSMC energy spectra of $^{11}$Be and $^{11}$B are shown in Figs.~\ref{fig:levels_be11} and~\ref{fig:levels_b11}, respectively. We present only states corresponding to experimentally bound states with respect to the $^{10}$Be$+$p (for $^{11}$B) and  $^{10}$Be$+$n (for $^{11}$Be) threshold.

The low-lying level ordering in $^{11}$Be is important because the
energies of the $1/2^+$ and $1/2^-$ states are inverted
compared to what would be expected from a standard shell-model picture. Our chosen interaction reproduces this parity inversion in the resulting NCSMC levels, as can be seen in Fig.~\ref{fig:levels_be11}.
%
In previous
NCSMC calculations using different interactions, the parity inversion could be
reproduced with the non-local N$^2$LO$_{\rm sat}$ interaction~\cite{N2LOsat} but not using interactions with local $3N$~\cite{Calci:2016}. Since the $NN$-N$^{4}$LO(500)$+3N_{\rm lnl}$ interaction reproduces this peculiarity of $^{11}$Be, we feel that (i) the non-locality of the $3N$ interaction is an important feature for the description of exotic nuclei (i.e. it leads to a better reproduction of the extended nuclear density) and (ii) the selected interaction is appropriate for calculating the $^{11}$Be $\beta$ decay. The present description is still showing discrepancies with data as the $1/2^+$ state is less bound than in experiment and the experimentally very weakly bound $1/2^-$ state is obtained just above the $^{10}$Be$+$n threshold.

The lowest negative-parity $^{11}$B levels are well reproduced in our calculations with an under-prediction of the splitting between the $3/2^-$ ground state and the $1/2^-$ first excited state, indicating a weaker spin-orbit strength of the employed interaction. The lowest $5/2^-$ and $7/2^-$ as well as $3/2^-_2$ and $5/2^-_2$ states match well the experimental energies. The experimental $3/2^-_3$ state at 8.56 MeV (not shown in Fig.~\ref{fig:levels_b11}) with a pronounced $\alpha$-cluster structure is, however, overpredicted in the present NCSMC calculations~\cite{Kawabata:2007}. Similarly, the lowest positive parity states appear more than 2.5 MeV too high. This is in part a consequence of the missing $^7$Li${+}\alpha$ mass partition in the calculations that we were not able to include for technical reasons. The $^7$Li${+}\alpha$ threshold appears experimentally 2.56 MeV below the $^{10}$Be$+$p threshold. The focus of this work is in particular on the $T{=}1/2$ $1/2^+$ and $3/2^+$ states in $^{11}$B. As seen in Fig.~\ref{fig:levels_b11}, we obtain one $T{=}1/2$ $1/2^+$ bound state and the lowest $3/2^+$ state is just above the  $^{10}$Be$+$p threshold.

\begin{figure}[h]
    \begin{center}
       \includegraphics[scale=1.13]{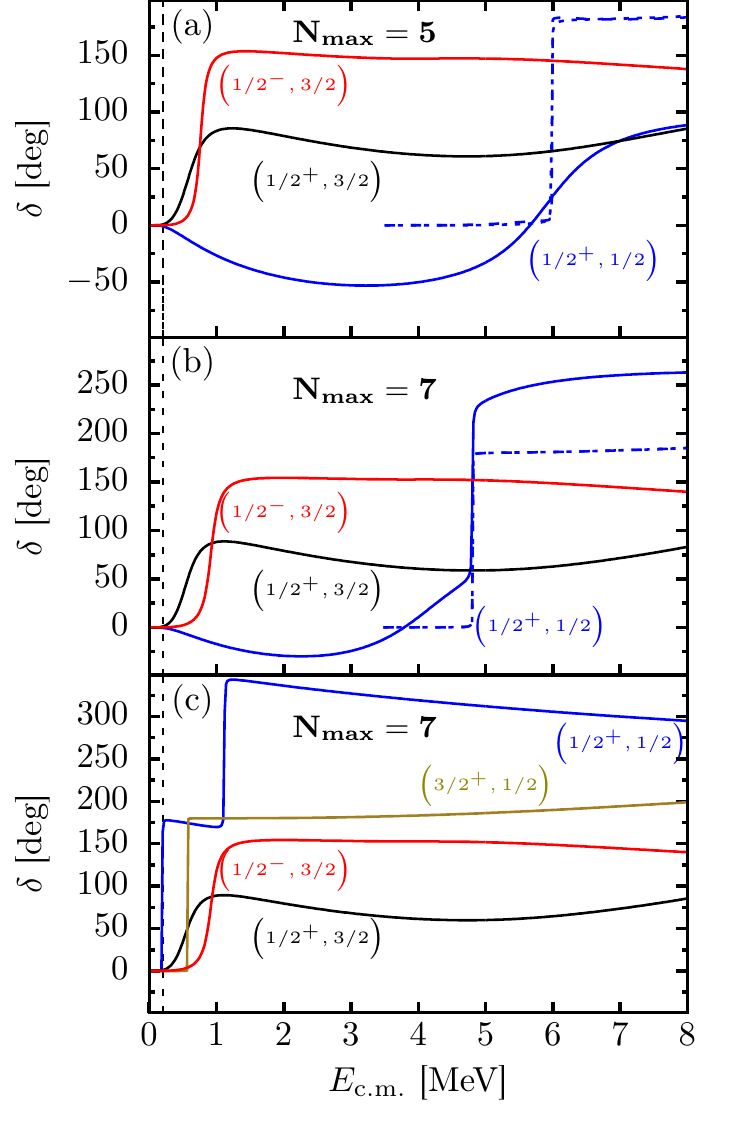}
    \end{center}
    \caption{NCSMC-calculated $^{10}$Be$+$p diagonal phase shifts. The solid lines correspond to the $^2S_{1/2}$ (or $^2P_{1/2}$) channels while the dashed and dotted curves correspond to the $^4D_{1/2}$ and $^6D_{1/2}$ channels, respectively. The vertical dashed line indicates the experimentally-predicted location of the $(1/2^+,1/2)$ resonance at 197 keV. (a) NCSMC-calculated phase shifts using $N_{\rm max}{=}5$ basis size. (b) NCSMC-calculated phase shifts using $N_{\rm max}{=}7$ basis size.  (c) Phenomenologically adjusted, NCSMC$_{\mathrm{pheno}}$, phase shifts such that the $(1/2^+,1/2)$ resonance coincides with the experimentally-predicted resonance at 197 keV. The $(3/2^+,1/2)$ resonance is the result of shifting the degenerate states to the corresponding experimental levels (see Fig.~\ref{fig:levels_b11}). See the text for details. $E_{\rm c.m.}$ is the kinetic energy of $^{10}$Be$+$p in the center-of-mass frame.}
 \label{fig:phase}
 \end{figure}
 \begin{figure}[h]
    \begin{center}
       \includegraphics[scale=1.13]{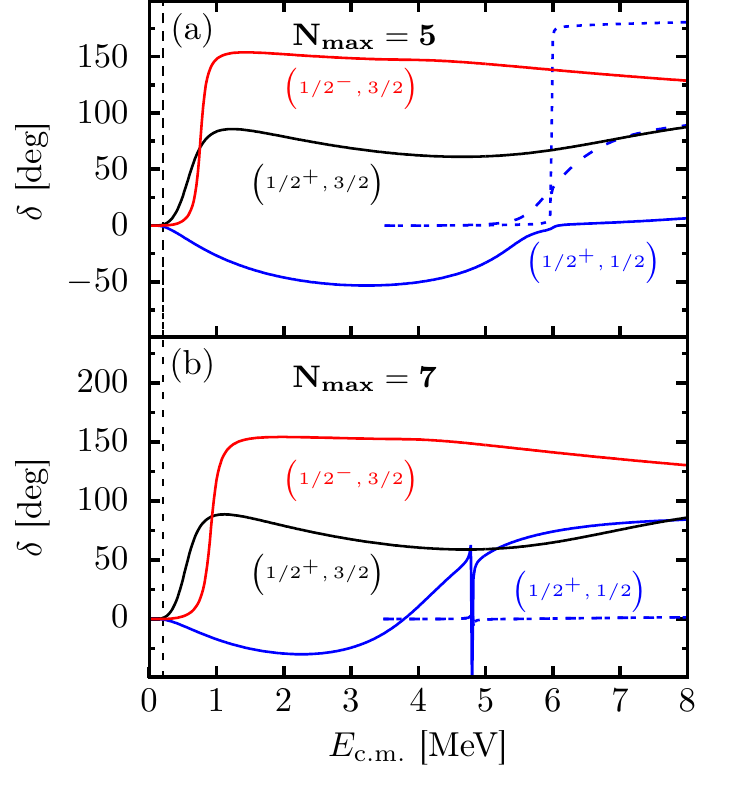}
    \end{center}
    \caption{NCSMC-calculated $^{10}$Be$+$p eigenphase shifts. The vertical dashed line indicates the experimentally-predicted location of the $(1/2^+,1/2)$ resonance at 197 keV. (a) NCSMC-calculated phase shifts using $N_{\mathrm{max}}=5$ basis size. (b) NCSMC-calculated phase shifts using $N_{\rm max}{=}7$ basis size. The solid lines correspond to the first while the dashed and dotted curves correspond to the second and the third eigenphase shift, respectively.
    The sharp resonance in the $(1/2^+,1/2)$ solid line in (b) is not shifted by $\pi$ for clarity. 
    $E_{\rm c.m.}$ is the kinetic energy of $^{10}$Be$+$p in the center-of-mass frame.}
 \label{fig:eigenphase}
\end{figure}
In addition to bound-state energy spectra, the NCSMC provides the low-energy phase shifts and eigenphase shifts for the $^{10}$Be$+$p channel, see Figs.~\ref{fig:phase} and~\ref{fig:eigenphase}, respectively. To show the effect of increasing the basis size, both the $N_{\rm max}{=}5$ and $N_{\rm max}{=}7$ results are presented in Fig.~\ref{fig:phase}(a) and
Fig.~\ref{fig:phase}(b), respectively, and similarly for the eigenphase shifts in Fig.~\ref{fig:eigenphase}. The  $(1/2^-,3/2)$ and
$(1/2^+,3/2)$ phase shifts, isobaric analogues of the two bound states in $^{11}$Be, are plotted to emphasize the parity inversion discussed in the previous paragraph.  Both the $N_{\rm max}{=}5$ and $N_{\rm max}{=}7$ calculations reproduce the parity inversion in that the
$(1/2^+,3/2)$ resonance is lower in energy than the $(1/2^-,3/2)$
resonance. The separation between these two resonances is more pronounced in the more converged result
of $N_{\rm max}{=}7$ (see Fig.~\ref{fig:phase}(b)).
 \begin{table}
    \centering
         {\renewcommand{\arraystretch}{1.15}
   \begin{tabular}{c c c c c} 
      \hline
      \multirow{2}{1.5cm}{$J^\pi$} & $S(^{11}$B$\rightarrow^{10}$Be$)$ & \multicolumn{2}{c}{$S(^{11}$B$\rightarrow^{10}$B$)$} & $S(^{11}$B$\rightarrow^{7}$Li$)$ \\
      & $(0^+,1)$ & $(1_1^+,0)$ & $(1_2^+,0)$ & $(3/2^+,1/2)$ \\
      \hline
      \hline
      $1/2_1^+$ & 0.276 & 0.250 & $2\times10^{-4}$ & 0.218 \\
      \hline
      $1/2_2^+$ & 0.0525 & 0.171 & 0.562 & 0.002 \\
      \hline
      $1/2_3^+$ & 0.067 & 0.231 & 0.188 & 0.011 \\
      \hline
      $3/2_1^+$ & 0.079 & $6\times10^{-4}$ & 0.215 & 0.009 \\
      \hline
      $3/2_2^+$ & $4\times10^{-4}$ & 0.581 & 0.002 & 0.012 \\
      \hline
      $3/2_3^+$ & $6\times10^{-4}$ & 0.011 & 0.006 & 0.021 \\
      \hline
      $3/2_4^+$ & 0.067 & 0.034 & 0.35 & 0.006 \\
      \hline
   \end{tabular}
}
   \caption{NCSM spectroscopic factors calculated from the overlap between select $^{11}$B states and $^{10}$Be, $^{10}$B, and $^{7}$Li states. The first column displays the $J^\pi$ of each considered $T=1/2$ $^{11}$B state. Spectroscopic factors of the ground states of $^{10}$B and $^{10}$Be and the first two $(1^+,0)$ excited states of $^{10}$B are presented. }
    \label{table:spec-fac}
 \end{table}

Shifting focus to the $(1/2^+,1/2)$ phase shifts, two resonances are present - one broad and one sharp. Thus, the NCSMC supports the existence of a $(1/2^+,1/2)$ resonance in the $^{10}$Be$+$p system.  However, the energy of either resonance is higher than the experimental prediction of 197 keV.  Furthermore, it is unclear which resonance should
    correspond to the experimental prediction. To gain insight in the structure of the $1/2^+$ states, we follow Ref.~\cite{PhysRevC.70.054324} to calculate the overlap
between the corresponding NCSM $^{11}$B states and the $^{10}$Be ground state,
$^{10}$B $(1^+,0)$ excited states, and the $^{7}$Li ground state (see
Table~\ref{table:spec-fac}). The first $1/2^+$ state with the largest overlap
with the $^{10}$Be ground state corresponds to the bound state shown in
Fig.~\ref{fig:levels_b11} while the second and third $1/2^+$ states,
corresponding to the two $(1/2^+,1/2)$ resonances in Fig.~\ref{fig:phase}, have
comparably low overlaps. While these proton spectroscopic factors do not
distinguish between the two resonances, the large $1/2^+_2$ $^{10}$B overlap
indicates that its corresponding NCSMC resonance contains significant
single-neutron content. From this, it is clear that the sharp resonance is
caused by the lack of $^{10}$B $1^+$ channels and would be broadened by their
inclusion in the NCSMC calculation. 
With the sharp resonance classified, the broad resonance must therefore correspond to the $1/2_3^+$ NCSM state and be the candidate for the experimentally-measured proton resonance.

\subsection{Phenomenologically adjusted NCSMC}

With the resonance identified,
the next step is to calculate the
branching ratio for the $\beta$-decay from the $^{11}$Be ground state to the $^{11}$B proton
resonance. In order to better evaluate how well this resonance explains the experimentally observed
branching ratio, we introduce a phenomenological shift to the NCSMC calculation (see, e.g., Ref.~\cite{Calci:2016}) resulting in the calculated resonance lying at 197 keV, see Fig.~\ref{fig:phase}(c).
This approach, dubbed NCSMC$_\mathrm{pheno}$, proceeds
by using the $^{10}$Be, $^{11}$B, and $^{11}$Be NCSM eigenenergies as adjustable parameters in the NCSMC equations. First, the $^{10}$Be $2^+$ excitation energy is set to its experimental value (a change from NCSM calculated 3.48 MeV to experimental 3.37 MeV). We then adjust only the $1/2^+$ and $3/2^+$ channels relevant for the $\beta$ decay calculations. Consequently, there is almost no change in energies of negative parity states and the $5/2^+$ state in Figs.~\ref{fig:levels_be11} and \ref{fig:levels_b11}, and in the $T{=}3/2$ phase shift in Fig.~\ref{fig:phase}(c).
The fact that the $(1/2^+,1/2)$ resonances shift slightly more than 1 MeV lower in
energy when increasing the basis size from $N_{\rm max}{=}5$ to the $N_{\rm max}{=}7$ (compare
Figs.~\ref{fig:phase}(a), \ref{fig:eigenphase}(a)  and ~\ref{fig:phase}(b), \ref{fig:eigenphase}(b)) implies that a larger model space can bring the resonance down further. Thus, this phenomenological shift is emulating
the effect of including more channels. In addition to shifting
the $(1/2^+,1/2)$ resonance to 197 keV, we shift the $^{11}$B $(3/2^+,1/2)$ levels as well as the $^{11}$Be ground state to
its experimental value, see Figs.~\ref{fig:levels_be11} and~\ref{fig:levels_b11}. These phenomenological shifts bring the NCSMC major shell splittings closer to experimental values.
A consequence of shifting the first two $(3/2^+,1/2)$ $^{11}$B levels is the emergence of a sharp resonance (previously broad and higher in energy) just 573 keV above threshold in Fig.~\ref{fig:phase}. The location of this resonance is similar to the predicted resonance at 262 keV in Ref.~\cite{Refsgaard:2019}. The authors of Ref.~\cite{Refsgaard:2019} suggest that this could be the unassigned resonance observed in Ref.~\cite{Kelley:2012}. 

\subsection{$^{11}$Be $\beta$ decay}\label{subsec:beta}

We first calculate $B(\mathrm{GT})$ for
the $\beta$-decay from the $^{11}$Be ground state to the first bound
$(1/2^+,1/2)$ state in $^{11}$B. This level is shifted to its
experimental value as well, see Fig.~\ref{fig:levels_b11}. The results are shown in
Table~\ref{table:bound_gt}. Also shown in Table~\ref{table:bound_gt} is the $B(\mathrm{GT})$
calculated for the $\beta$-decay to the three bound $(3/2^+,1/2)$
states. The third bound $(3/2^+,1/2)$ state is a result of phenomenologically shifting the first two levels to their corresponding experimental values.
To determine the half-life from $B(\mathrm{GT})$, we use~\cite{Bohr-Mottelson}: 

\begin{equation}
   fT^p_{1/2} = \frac{6141}{g_{A}^2B(\mathrm{GT})},
   \label{eq:logft}
\end{equation}

where $g_{A}{=}-1.27$ is the GT coupling constant, $f$ is a phase-space factor determined by the $Q$-value of the decay, and $T^p_{1/2}$ is the half-life of the decay. The half-life of the decay can be used to calculate the corresponding branching ratio using the following expression:
\begin{equation}
   b_p = \frac{T^{^{11}{\rm Be}}_{1/2}}{T^p_{1/2}},
   \label{eq:bp}
\end{equation}
where $T^{^{11}{\rm Be}}_{1/2} = 13.8\; s$ is the half-life of the $^{11}$Be nucleus. Using Eqs.~\eqref{eq:logft} and~\eqref{eq:bp}, the experimental branching ratios (see Ref.~\cite{Millener:1982}) were converted to the $B(\mathrm{GT})$ values in Table~\ref{table:bound_gt}.
It is clear from Table.~\ref{table:bound_gt} that the inclusion of the proton
and neutron channels in the NCSMC improves the calculated $B(\mathrm{GT})$ values over the NCSM results.
\begin{table}[h]
    \centering
         {\renewcommand{\arraystretch}{1.15}
   \begin{tabular}{cccc} 
      \hline
      $B(\mathrm{GT})$ & NCSM & NCSMC$_{\mathrm{pheno}}$ & Expt. \\ 
      \hline
      \hline
      $1/2^+_1$ &   0.341 & 0.277  & 0.004 \\
      \hline
      $3/2^+_1$ &   0.023 & 0.002  & 0.010 \\
      \hline
      $3/2^+_2$ &  2.92 & 0.286  & 0.228 \\
      \hline
      $3/2^+_3$ &   0.011 & $7\times10^{-5}$  & - \\
      \hline
   \end{tabular}
}
\caption{$B(\mathrm{GT})$ values from the ground state of $^{11}$Be to the specified bound states of $^{11}$B. The first column displays the $J^\pi$ of each considered $T=1/2$ $^{11}$B state. The NCSMC$_{\mathrm{pheno}}$ approach was applied. Both the NCSM and NCSMC$_{\mathrm{pheno}}$ values were obtained at $N_{\mathrm{max}}=7$. Experimental values obtained as branching ratios from Ref.~\cite{Millener:1982} and converted to $B(\mathrm{GT})$ values (see text).}
   \label{table:bound_gt}
 \end{table}
The transition strength to the $1/2_1^+$ state is unexpectedly large compared to the experimental strength. This is indicative of the mixing of strength between the bound $1/2_1^+$ state and the resonance in question.
It is also worth noting that the overlap between the $1/2_1^+$ state and both the $^7$Li$+\alpha$ ground and $^{10}$B$+$n $(1_1^+,0)$ states is significant (see Table~\ref{table:spec-fac}). Thus, the inclusion of these channels in the NCSMC calculation could address this discrepancy.
The NCSMC$_\mathrm{pheno}$ calculations of the experimentally strong transition to the $3/2^+_2$ state show great improvement over the NCSM calculations. The cause of the enhanced NCSM $B(\mathrm{GT})$ value can be attributed to the large overlap with the $^{10}$B$+$n $(1_1,0)$ state (see Table.~\ref{table:spec-fac}).
With the inclusion of the $^{10}$Be+$n$ channel in the NCSMC calculation, the halo structure of $^{11}$Be is reproduced which has a small overlap with the $^{11}$B $3/2^+_2$ state thus suppressing the NCSMC $B(\mathrm{GT})$.
\begin{figure}[h]
   \begin{center}
      \includegraphics[scale=1.13]{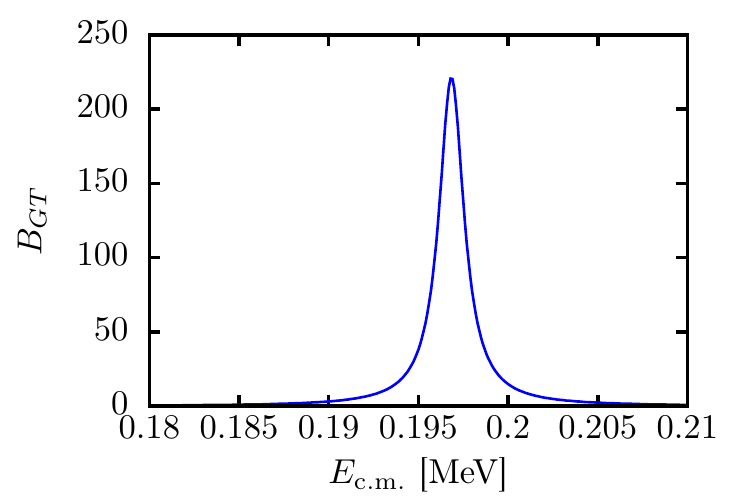}
   \end{center}
   \caption{Energy-dependent $B_{\mathrm{GT}}(E)$ calculated from the ground state of $^{11}$Be to the
   $(1/2^+,1/2)$ proton continuum of $^{11}$B using the NCSMC.}
   \label{fig:bgt}
 \end{figure}
 
The calculation of the branching ratio for the $\beta$-decay to the
$(1/2^+,1/2)$ resonance is more involved since the final state in Eq.~\eqref{eq:bgt}
is an energy-dependent scattering wave function. The scattering wave function is calculated in
$k$-space (using the same same convention from Ref.~\cite{Navratil:2016}), thus the matrix element in Eq.~\eqref{eq:bgt} must be evaluated with a volume integral in $k$-space,
\begin{equation}
   B(\mathrm{GT}) = 4\pi \int_{k_{min}}^{k_{max}} dkk^2B_{\mathrm{GT}}(k),
   \label{eq:bgt_kspace}
\end{equation}
where $k_{min}$ and $k_{max}$ are chosen to isolate the location of the resonance. 
In order to calculate the branching ratio for this process, $B_{\mathrm{GT}}(k)$ must be folded with the energy-dependent phase-space factor $f$.
Thus, we first transform Eq.~\eqref{eq:bgt_kspace} to energy-space such that 
\begin{equation}
   B(GT) = \int_{E_{min}}^{E_{max}} dEB_{\mathrm{GT}}(E),
   \label{eq:bgt_espace}
\end{equation}
where
\begin{equation}
   B_{\mathrm{GT}}(E) := \frac{4\pi\mu k}{\hbar^2}B_{\mathrm{GT}}(k)
   \label{eq:kspace_transform}
\end{equation}
and $k=\frac{\sqrt{2\mu E}}{\hbar}$ with $\mu$ the reduced mass.
The energy-dependent $B_{\mathrm{GT}}(E)$ is shown in Fig.~\ref{fig:bgt}, where it displays a prominent peak at the resonance energy of 197 keV.  
The sharp peak in $B_{\mathrm{GT}}(E)$ is a visual representation of how a resonance can enhance the $\beta$-decay transition rate to the continuum.
The $Q$-value at a given energy can be expressed as $Q = Q_0-E$ where $Q_0=0.281$~MeV is the Q-value evaluated at the p${+}^{10}$Be threshold. The branching ratio of the $\beta$-decay to the resonance can be calculated by combining Eqs.~\eqref{eq:logft},~\eqref{eq:bp}, 
and~\eqref{eq:bgt_espace} in the following way: 
\begin{equation}
   b_p = \frac{T_{1/2}^{^{11}{\rm Be}}g_{A}^2}{6141}\int_{E_{min}}^{E_{max}} dEB_{\mathrm{GT}}(E)f(Q_0-E).
   \label{eq:bp_e}
\end{equation}

 \begin{widetext}

     \begin{table}[h]
         {\renewcommand{\arraystretch}{1.15}
   \begin{tabular}{cccccc} 
      \hline
      \multirow{2}{1.7cm}{$(1/2^+,1/2)$} & \multicolumn{2}{c}{$N_{\rm max}=5$} & \multicolumn{2}{c}{$N_{\rm max}=7$} & \multirow{2}{0.5cm}{Expt.} \\
      & NCSM & NCSMC$_{\mathrm{pheno}}$ & NCSM & NCSMC$_{\mathrm{pheno}}$ \\ 
      \hline
      \hline
      $B(\mathrm{GT})$ & 1.95 & 0.325 &  1.39 & 0.565  & $5.5^{8.3}_{3.3}$ \\
      \hline
      $b_p$ & - & $7.4\times 10^{-7}$ & - & $1.3\times 10^{-6}$  & $1.3(3)\times 10^{-5}$ \\
      \hline
   \end{tabular}
}
\caption{$B(\mathrm{GT})$ $\beta$-decay values for the transition from the ground state of $^{11}$Be to the p+$^{10}$Be resonance. The NCSM results correspond to the $(1/2_3^+,1/2)$ state. The experimental values are those presented in Ref.~\cite{Ayyad:2019}}
   \label{table:bgt_res}
\end{table}

\end{widetext}

Using Eq.~\eqref{eq:bp_e}, we integrate $B_{\mathrm{GT}}(E)$ over the energy-range shown in Fig.~\ref{fig:bgt} resulting in the values shown in Table~\ref{table:bgt_res}. The NCSMC $B(\mathrm{GT})$ value reported in Table~\ref{table:bgt_res} is calculated as $B(\mathrm{GT}) = \int dEB_{\mathrm{GT}}(E)$. 
Although the calculated branching ratio, $b_p = (1.3\pm0.5)\times 10^{-6}$, is lower than the two experimental observations~\cite{Riisager:2014,Ayyad:2019}, it is consistent when taking into account that the non-resonant decay branching ratio calculated in Ref.~\cite{Baye:2011} is two orders of magnitude smaller.
The uncertainty is estimated by comparing the NCSMC$_\mathrm{pheno}$ results using $N_{\rm max}=5$ and $N_{\rm max}=7$.
To verify that we shifted the correct resonance, we also calculated $B(\mathrm{GT}) = 0.0483$ for the transition to the $(1/2^+,1/2)$ resonance located around $1$ MeV in Fig.~\ref{fig:phase}(c). This weak transition strength confirms that we have chosen the correct candidate. We also investigate the sharp $(3/2^+,1/2)$ resonance in Fig.~\ref{fig:phase}(c) as it was suggested as a candidate for the large branching ratio~\cite{Ayyad:2019}. While there is a non-negligible overlap between $3/2^+_4$ and $^{10}$Be+p (see Table~\ref{table:spec-fac}), we calculate that $B(\mathrm{GT})=0.00131$ for this particular resonance. This is contrary to the large $B(\mathrm{GT})$ empirically calculated for this resonance in Ref~\cite{Refsgaard:2019}. 
It is also worth nothing that the $\alpha$ spectroscopic factors for the $1/2_2$ and $1/2_3$ states in Table~\ref{table:spec-fac} are significantly smaller than those calculated in the shell model calculations in Ref.~\cite{Volya:2020}. The smaller spectroscopic factors indicate that the proton decay channel will not have as much competition with the $\alpha$ decay channel as was indicated of Ref.~\cite{Volya:2020}. 
With these considerations, we conclude that the large observed branching is due to the existence of this $(1/2^+,1/2)$ p+$^{10}$Be resonance.

\section{Conclusions}\label{concl}

Using the \textit{ab initio} NCSMC, we investigated the bound energy-levels and low-lying
resonances in both $^{11}$Be and $^{11}$B. The two- and three-body
interactions, $NN$-N$^{4}$LO(500)$+3N_{\rm lnl}$, produce realistic energy
spectra in both $^{11}$Be and $^{11}$B. The experimentally-observed parity
inversion in $^{11}$Be is reproduced. We
identified two $(1/2^+,1/2)$ resonances in the p+$^{10}$Be continuum and
determined that the broad resonance dominated by the $1/2^+_3$ NCSM
state is the candidate resonance. The location of
the resonance is several MeV higher than the experimentally predicted location
of 197 keV although its position decreases dramatically with the increasing basis size.  In order to determine if this resonance can explain the large
branching ratio observed by experiment, we phenomenologically shifted the NCSMC
resonance to 197 keV.

With the resonance determined, we developed the procedure to calculate $B(\mathrm{GT})$
fully within the NCSMC. Appendix~\ref{sec:appendix} contains
the relevant derivations. We then extended the calculation to the continuum by
deriving the necessary expressions to calculate the branching ratio for the
decay to a resonance state (in the p+$^{10}$Be continuum). The resulting branching ratio for the $\beta$-decay to the 
candidate $(1/2^+,1/2)$ resonance is $b_p = (1.3\pm0.5)\times10^{-6}$.
This calculated branching ratio is consistent with both experimental branching
ratios reported Ref.~\cite{Riisager:2014} and Ref.~\cite{Ayyad:2019} taking into account that the non-resonant decay branching ratio is two orders of magnitude smaller. Furthermore, we calculate a small $B(\mathrm{GT})$ for the nearby $(3/2^+,1/2)$ resonance which further reinforces our conclusions.

The present calculations can be further improved by including the $^7$Li+$\alpha$ and $^{10}$B+n mass partitions. The former would particularly impact the description of the $1/2^+_1$ bound state and presumably redistribute some of the $B(\mathrm{GT})$ strength from the bound state to the $(1/2^+,1/2)$ resonance. 
Work in this direction is under way, see Ref.~\cite{Kravvaris:2020cvn} for the latest developments regarding the inclusion of $\alpha$ clustering in the NCSMC. 

\section{Acknowledgements}
This work was supported by the NSERC Grant No. SAPIN-2016-00033 and by the U.S. Department of Energy, Office of Science, Office of Nuclear Physics, under Work Proposals No. SCW0498. TRIUMF receives federal funding via a contribution agreement with the National Research Council of Canada. This work was prepared in part by LLNL under Contract No. DE-AC52-07NA27344. Computing support came from an INCITE Award on the Summit supercomputer of the Oak Ridge Leadership Computing Facility (OLCF) at ORNL, from Livermore Computing, Westgrid and Compute Canada.

\appendix
\section{Derivation of $B(\mathrm{GT})$ in the NCSMC}
\label{sec:appendix}

To calculate $B(\mathrm{GT})$, the reduced matrix element of the GT operator is calculated with NCSMC initial and final states. We derive this reduced matrix element for a general coordinate-independent one-body operator, $\mathcal{O}^{(\kappa\tau)}_{\mu_\tau}$ with $\kappa$ and $\tau$ the spin and isospin rank, respectively, and $\mu_\tau$ the isospin projection, e.g., GT, Fermi, spin part of the M1 operator. The GT operator corresponds to the case where $\kappa{=}1$, $\tau{=}1$, and $\mu_\tau{=}+1$.
Inserting the form of the NCSMC wave function (Eqs.\eqref{eq:ncsmc} and \eqref{eq:Be10N_rgm_state}) in Eq.~\eqref{eq:bgt} leads to four contributions:
\begin{align}
   &\Braket{\Psi^{J_f^{\pi_f} T_f}_{f \;\; M_{T_f}}||\hat{\mathcal{O}}^{(\kappa\tau)}_{\mu_\tau}||\Psi^{J_i^{\pi_i} T_i}_{i \;\;M_{T_i}}} = \nonumber\\
   &\sum_{\lambda_i\lambda_f} \tilde{c}^*_{\lambda_f} \tilde{c}_{\lambda_i} \Braket{A\lambda_f J_fT_fM_{T_f}||\hat{\mathcal{O}}^{(\kappa\tau)}_{\mu_\tau}||A\lambda_i J_iT_iM_{T_i}} \nonumber\\
   &+\sum_{\lambda_f\nu_i} \tilde{c}^*_{\lambda_f} \int_0^\infty dr r^2 \tilde{\chi}_{\nu_i}(r)\textstyle{\frac{1}{r}} \nonumber\\
  &\times \Braket{A\lambda_f J_fT_fM_{T_f}||\hat{\mathcal{O}}^{(\kappa\tau)}_{\mu_\tau}\hat{\mathcal{A}}_{\nu_i}||\Phi_{\nu_i r}^{J_iT_iM_{T_i}}}\nonumber \\
   &+\sum_{\lambda_i\nu_f} \tilde{c}_{\lambda_i} \int_0^\infty dr r^2 \tilde{\chi}^*_{\nu_f}(r)\textstyle{\frac{1}{r}} \nonumber\\
  &\times\Braket{\Phi_{\nu_f r}^{J_fT_fM_{T_f}}||\hat{\mathcal{A}}_{\nu_f}\hat{\mathcal{O}}^{(\kappa\tau)}_{\mu_\tau}||A\lambda_i J_iT_iM_{T_i}} \nonumber\\
   &+\sum_{\nu_f\nu_i} \int_0^\infty dr' r'^2 \tilde{\chi}^*_{\nu_f}(r') \textstyle{\frac{1}{r'}} \int_0^\infty dr r^2 \tilde{\chi}_{\nu_i}(r) \textstyle{\frac{1}{r}}
     \nonumber \\
  &\times\Braket{\Phi_{\nu_f r'}^{J_fT_fM_{T_f}}||\hat{\mathcal{A}}_{\nu_f}\hat{\mathcal{O}}^{(\kappa\tau)}_{\mu_\tau}\hat{\mathcal{A}}_{\nu_i}||\Phi_{\nu_i r}^{J_iT_iM_{T_i}}} \; .
   \label{eq:four-terms}
\end{align}
The expansion coefficients $\tilde{c}$ and $\tilde{\chi}(r)$ can be related to the expansion (\ref{eq:ncsmc}) and NCSMC norm kernels according to Eqs. (20), (32) and (33) in Ref.~\cite{Simone:2013}. We note that the $J^\pi T M_T$ quantum number dependence of these coefficients is not displayed to simplify the notation. Similarly, we omit the parity quantum number in most of the states. In the present case, $A{=}11$. We also note that matrix element (\ref{eq:four-terms}) is reduced only in spin-space, not isospin-space.

In order to proceed with the derivation, all contributions from localized terms are calculated in a HO basis and then converted to coordinate space with the following relation~\cite{Quaglioni:2009}:
\begin{align}
   \Ket{\Phi_{\nu r}^{JTM_{T}}} = \sum_n \Ket{\Phi_{\nu n}^{JTM_{T}}} R_{n\ell}(r),
   \label{eq:HO}
\end{align}
where $R_{nl}(r)$ are radial HO functions and
\begin{align}
    \ket{\Phi^{JTM_T}_{\nu n}} = &\Big[ \big( \ket{^{10} {\rm Be} \, \alpha_1 I_1^{\pi_1} T_1} \ket{N \, \tfrac12^{\texttt{+}}\tfrac12} \big)^{(sT)} \nonumber \\
     &\times\, Y_\ell(\hat{r}_{10,1}) \Big]^{(J^{\pi}T)}_{M_T} 
    \,R_{n\ell}(r_{10,1}) \; ,
    \label{eq:Be10N_HO_state}
\end{align}
with $N$ representing p for $^{11}$B and n for $^{11}$Be. See also the discussion in Sect. II.C.1 in Ref.~\cite{Quaglioni:2009}.

The process to calculate each of the terms in Eq.~\eqref{eq:four-terms} is analogous to the procedure for calculating the norm and Hamiltonian kernels for the NCSMC~\cite{Navratil:2016}.
The first term in Eq.~\eqref{eq:four-terms} involves NCSM matrix elements of $\hat{\mathcal{O}}^{(\kappa\tau)}_{\mu_\tau}$ calculated using standard second-quantization techniques. The second and third terms are analogous to the coupling kernels when solving the NCSMC Hamiltonian equations~\cite{Navratil:2016}. In order to continue, we represent the operator in a second-quantized form:
\begin{align}
   \hat{\mathcal{O}}^{(\kappa\tau)}_{\mu_\kappa\mu_\tau} =
   \left(\frac{-1}{\hat{\kappa}}\right)\sum\Braket{j_a\textstyle{\frac{1}{2}}m_{t_a}||\hat{\mathcal{O}}^{(\kappa\tau)}_{\mu_\tau}||j_b\textstyle{\frac{1}{2}}m_{t_b}}
   \nonumber \\
   \times\hat{a}^\dagger_{j_am_a\textstyle{\frac{1}{2}}m_{t_a}}\tilde{a}_{j_bm_b\textstyle{\frac{1}{2}}m_{t_b}}
   \Braket{j_am_aj_bm_b|\kappa\mu_\kappa} \; ,
   \label{eq:GT}
\end{align}
with $\tilde{a}_{j m \textstyle{\frac{1}{2}} m_{t}}{=}(-1)^{j-m} a_{j\; -m\; \textstyle{\frac{1}{2}} m_{t}}$. To expand this term, we also represent the cluster wave function in a second-quantized form (see Sect. II.D.2 of Ref.~\cite{Quaglioni:2009} and Sect. II.A of Ref.~\cite{PhysRevC.88.054622}):
\begin{align}
   \hat{\mathcal{A}}_{\nu} & \ket{\Phi_{\nu n}^{J_iM_iT_iM_{T_i}}}_{\rm SD} = \sum 
   (-1)^{I_1+J_i+j}\hat{s}\hat{j}
   \begin{Bmatrix}
      I_1 & \textstyle{\frac{1}{2}} & s \\
      \ell & J_i & j
   \end{Bmatrix}
\nonumber\\
&\times
   \Braket{I_1M_1jm|J_iM_i}\Braket{T_1M_{T_1}\textstyle{\frac{1}{2}}m_t|T_iM_{T_i}}
\nonumber\\
&\times
   \hat{a}^\dagger_{n\ell jm\textstyle{\frac{1}{2}}m_t}\ket{^{10}{\rm Be}\,\alpha_1I_1M_1T_1M_{T_1}}_{\rm SD},
   \label{eq:phi}
\end{align}
where $\ket{^{10}{\rm Be}\,\alpha_1I_1M_1T_1M_{T_1}}_{\rm SD}$ is an NCSM state of the $(A{-}1)$ system $^{10}$Be expanded in the HO Slater Determinant (SD) basis, 
$
   \begin{Bmatrix}
      I_1 & I_2 & s \\
      \ell & J_i & j
   \end{Bmatrix}
$ 
is the standard Wigner 6j-symbol~\cite{Gottfried}, and $\hat{x}$ denotes $\sqrt{2x+1}$. The $nl$ HO quantum numbers in the $nlj$ single-particle states are omitted from now on for simplicity. The relationship between the SD eigenstates entering Eq.~\eqref{eq:phi} and the relative coordinate eigenstates in Eqs.~\eqref{eq:Be10N_rgm_state} and \eqref{eq:Be10N_HO_state} is given by
\begin{align}
   \ket{^{10}{\rm Be}\,\alpha_1I_1M_1T_1M_{T_1}}_{\rm SD}&=\ket{^{10}{\rm Be}\,\alpha_1I_1M_1T_1M_{T_1}} \nonumber \\
   &\times\, \varphi_{00}(\vec{R}^{(10)}_{\rm c.m.}) \;,
   \label{eq:10Becm}
\end{align}
with $\vec{R}^{(10)}_{\rm c.m.}$ the c.m. coordinate of $^{10}$Be and $\varphi_{00}$ the HO wave function of the c.m. motion.
Using Eq.~\eqref{eq:phi}, the coupling term becomes:
\begin{widetext}
\begin{align}
   &_{\rm SD}\Braket{A\lambda J_fT_fM_{T_f}||\hat{\mathcal{O}}^{(\kappa\tau)}_{\mu_\tau}\hat{\mathcal{A}}_\nu||\Phi_{\nu n}^{J_iT_iM_{T_i}}}_{\rm SD} = 
\sum (-1)^{1+\kappa+J_f-J_i}\hat{J}_i\hat{s}\hat{j}
   \begin{Bmatrix}
      I_1 & j & J_i \\
      \kappa & J_f & j_a
   \end{Bmatrix}
   \begin{Bmatrix} 
      I_1 & \textstyle{\frac{1}{2}} & s \\
      \ell & J_i & j
   \end{Bmatrix}\Braket{T_1M_{T_1}\textstyle{\frac{1}{2}}m_t|T_iM_{T_i}} \nonumber \\
   &\times\Braket{j_a\textstyle{\frac{1}{2}}m_{t_a}||\hat{\mathcal{O}}^{(\kappa\tau)}_{\mu_\tau}||j\textstyle{\frac{1}{2}}m_t}\,_{\rm SD}\Braket{A\lambda J_fT_fM_{T_f}||\hat{a}^\dagger_{j_a\textstyle{\frac{1}{2}}m_{t_a}}||A{-}1\, \alpha_1I_1T_1M_{T_1}}_{\rm SD} \nonumber \\
   &+ \sum (-1)^{J_f-J_1+J_2+j_b+\kappa-J_i}\hat{s}\hat{j}\hat{J}_i\hat{J}_1\hat{J}_2
   \begin{Bmatrix}
      j_a & j_b & \kappa \\
      J_1 & j & J_2
   \end{Bmatrix}
   \begin{Bmatrix}
      I_1 & j & J_i \\
      \kappa & J_f & J_1
   \end{Bmatrix}
   \begin{Bmatrix} 
      I_1 & \textstyle{\frac{1}{2}} & s \\
      \ell & J_i & j
   \end{Bmatrix}\Braket{T_1M_{T_1}\textstyle{\frac{1}{2}}m_t|T_iM_{T_i}} \nonumber \\
   &\times 
   \Braket{j_a\textstyle{\frac{1}{2}}m_{t_a}||\hat{\mathcal{O}}^{(\kappa\tau)}_{\mu_\tau}||j_b\textstyle{\frac{1}{2}}m_{t_b}}
   \,_{\rm SD}\Braket{A\lambda
      J_fT_fM_{T_f}||\left[(\hat{a}^\dagger_{j_a\textstyle{\frac{1}{2}}m_{t_a}}\hat{a}^\dagger_{j\textstyle{\frac{1}{2}}m_t})^{(J_2)}
      \tilde{a}_{j_b\textstyle{\frac{1}{2}}m_{t_b}}\right]^{(J_1)}||A{-}1\,\alpha_1I_1T_1M_{T_1}}_{\rm SD},
      \label{eq:coupling-kernel}
\end{align}
where $\Braket{\hat{a}^\dagger_{j_a\textstyle{\frac{1}{2}}m_{t_a}}}$ and 
$\Braket{\left[\left(\hat{a}^\dagger_{j_a\textstyle{\frac{1}{2}}m_{t_a}}\hat{a}^\dagger_{j\textstyle{\frac{1}{2}}m_t}\right)^{(J_2)}\tilde{a}_{j_b\textstyle{\frac{1}{2}}m_{t_b}}\right]^{(J_1)}}$ 
are 1-particle and 2-particle-1-hole transition matrices between the composite and the target (here $^{11}$B and $^{10}$Be) NCSM wave functions. 
Similarly as for the target eigenstates \eqref{eq:10Becm}, the composite system ($^{11}$Be or $^{11}$B) eigenstates expanded in the HO SD basis are related to the relative coordinate eigenstates appearing in Eq.~\eqref{eq:four-terms} by
%
\begin{equation}
   \Ket{A\lambda J T M_{T}}_{\rm SD} = \Ket{A\lambda J T M_{T}} \varphi_{00}(\vec{\xi}_0) 
   \label{eq:A_s=cm} \; ,
\end{equation}
with $\vec{\xi}_0$ the c.m. coordinate of the composite $A$-nucleon system and $\varphi_{00}$ the HO wave function of the c.m. motion. See also the appendix of Ref.~\cite{Simone:2013} and Ref.~\cite{PhysRevC.88.054622}.
The third term in Eq.~\eqref{eq:four-terms} is calculated by taking advantage of the following relation:
\begin{align}
   &_{\rm SD}\Braket{A\lambda_fJ_fT_fM_{T_f}||
   \hat{\mathcal{O}}^{(\kappa\tau)}_{\mu_\tau}\hat{\mathcal{A}}_\nu||\Phi_{\nu n}^{J_iT_iM_{T_i}}}_{\rm SD}= (-1)^{J_f-J_i} 
   \,\; _{\rm SD}\Braket{\Phi_{\nu n}^{J_iT_iM_{T_i}}||\left[\hat{\mathcal{O}}^{(\kappa\tau)}_{\mu_\tau}\hat{\mathcal{A}}_{\nu}\right]^\dagger||A\lambda_f J_fT_fM_{T_f}}_{\rm SD}. 
   \label{eq:transpose}
\end{align}
Thus, Eq.~\eqref{eq:coupling-kernel} is sufficient for both coupling terms in Eq.~\eqref{eq:four-terms}.

The last term in Eq.~\eqref{eq:four-terms} is derived using a similar procedure:
%
\begin{align}
   &_{\rm SD}\Braket{\Phi_{\nu' n'}^{J_fT_fM_{T_f}}||\hat{A}_{\nu'}\hat{\mathcal{O}}^{(\kappa\tau)}_{\mu_\tau}\hat{A}_{\nu}||\Phi_{\nu n}^{J_iT_iM_{T_i}}}_{\rm SD} = 
   \delta_{\alpha_1\alpha_1'}\delta_{I_1I_1'}\delta_{T_1T_1'}\delta_{M_{T_1}M'_{T_1}}\hat{J_f}\hat{J_i}\hat{s}\hat{s}' 
\nonumber \\
& \times
   \sum\hat{j}\hat{j}'(-1)^{I_1+\kappa+J_i+j'}
\begin{Bmatrix}
    I_1 & j & J_i \\
    \kappa & J_f & j' \\
\end{Bmatrix}
\begin{Bmatrix}
    I_1 & \textstyle{\frac{1}{2}} & s \\
    \ell & J_i & j \\
\end{Bmatrix}
\begin{Bmatrix}
    I_1 & \textstyle{\frac{1}{2}} & s' \\
    \ell' & J_f & j' \\
\end{Bmatrix}
\nonumber\\
& \times
\Braket{j'\textstyle{\frac{1}{2}}m_t'||\hat{\mathcal{O}}^{(\kappa\tau)}_{\mu_\tau}||j\textstyle{\frac{1}{2}}m_t}
\Braket{T_1M_{T_1}\textstyle{\frac{1}{2}}m_t|T_iM_{T_i}}\Braket{T_1M_{T_1}\textstyle{\frac{1}{2}}m_t'|T_fM_{T_f}}\nonumber\\
&-
\sum \hat{s}\hat{s}'\hat{j}\hat{j'}\hat{J_f}\hat{J_i}\hat{J_2}\hat{J_3}\hat{J_4}(-1)^{j'+j_a+\kappa+J_3+j_b+I_1-J_f}
\begin{Bmatrix}
   I_1 & \frac{1}{2} & s \\
    \ell & J_i & j \\
\end{Bmatrix}
\begin{Bmatrix}
    I_1' & \frac{1}{2} & s' \\
    \ell' & J_f & j' \\
\end{Bmatrix}
\begin{Bmatrix}
   - & j & I_1 & J_i \\
   j' & - & I_1' & J_f \\
   j_b & j_a & - & \kappa \\
   J_4 & J_2 & J_3 & -
\end{Bmatrix}
   \nonumber \\
   &\times
   _{\rm SD}\Braket{A{-}1\, \alpha_1'I_1'T_1'M_{T_1}'||\left[\left(\hat{a}^\dagger_{j_a\textstyle{\frac{1}{2}}m_{t_a}}\hat{a}^\dagger_{j\textstyle{\frac{1}{2}}m_t}\right)^{(J_2)}\left(\tilde{a}_{j'\textstyle{\frac{1}{2}}m_t'}\tilde{a}_{j_b\textstyle{\frac{1}{2}}m_{t_b}}\right)^{(J_4)}\right]^{(J_3)}||A{-}1\, \alpha_1I_1T_1M_{T_1}}_{\rm SD}
   \nonumber \\
&\times
\Braket{j_a\textstyle{\frac{1}{2}}m_{t_a}||\hat{\mathcal{O}}^{(\kappa\tau)}_{\mu_\tau}||j_b\textstyle{\frac{1}{2}}m_{t_b}}
\Braket{T_1'M_{T_1}'\textstyle{\frac{1}{2}}m_t'|T_fM_{T_f}}
   \Braket{T_1M_{T_1}\textstyle{\frac{1}{2}}m_t|T_iM_{T_i}}
   \nonumber\\ &-
   \sum \hat{s}\hat{s}'\hat{j}\hat{j}'\hat{J_f}\hat{J_i}\hat{J_2}(-1)^{j_a+j'+I_1+I_1'+\kappa+J_i+J_f}
\begin{Bmatrix}
    I_1 & \frac{1}{2} & s \\
    \ell & J_i & j \\
\end{Bmatrix}
\begin{Bmatrix}
    I_1' & \frac{1}{2} & s' \\
    \ell' & J_f & j' \\
\end{Bmatrix}
\begin{Bmatrix}
    j_a & j & \kappa \\
    J_i & J_f & I_1 \\
\end{Bmatrix}
\begin{Bmatrix}
    I_1 & j_a & J_f \\
    j' & I_1' & J_2 \\
\end{Bmatrix}
\nonumber \\
&\times\Braket{j_a\textstyle{\frac{1}{2}}m_{t_a}||\hat{\mathcal{O}}^{(\kappa\tau)}_{\mu_\tau}||j\textstyle{\frac{1}{2}}m_t}
_{\rm SD}\Braket{A{-}1\, \alpha_1'I_1'T_1'M_{T_1}'||\left(\hat{a}^\dagger_{j_a\textstyle{\frac{1}{2}}m_{t_a}}\tilde{a}_{j'\textstyle{\frac{1}{2}}m_t'}\right)^{(J_2)}||A{-}1\, \alpha_1I_1T_1M_{T_1}}_{\rm SD}
\nonumber \\
&\times
   \Braket{T_1M_{T_1}tm|T_iM_{T_i}}\Braket{T_1'M_{T_1}'\textstyle{\frac{1}{2}}m_t'|T_fM_{T_f}}
   \nonumber\\ &- 
   \sum \hat{j}\hat{j}'\hat{s}\hat{s}'\hat{J_f}\hat{J_i}\hat{J_1}(-1)^{j_b-j'-\kappa+1}
\begin{Bmatrix}
    I_1 & \frac{1}{2} & s \\
    \ell & J_i & j \\
\end{Bmatrix}
\begin{Bmatrix}
    I_1' & \frac{1}{2} & s' \\
    \ell' & J_f & j' \\
\end{Bmatrix}
\begin{Bmatrix}
    I_1 & j & J_i \\
    j_b & I_1' & J_1 \\
\end{Bmatrix}
\begin{Bmatrix}
    j' & j_b & \kappa \\
    J_i & J_f & I_1' \\
\end{Bmatrix}
\nonumber \\
   &\times
\Braket{j'\textstyle{\frac{1}{2}}m_{t'}||\hat{\mathcal{O}}^{(\kappa\tau)}_{\mu_\tau}||j_b\textstyle{\frac{1}{2}}m_{t_b}}  
   _{\rm SD}\Braket{A{-}1\, \alpha_1'I_1'T_1'M_{T_1}'||\left(\hat{a}^\dagger_{j\textstyle{\frac{1}{2}}m_t}\tilde{a}_{j_b\textstyle{\frac{1}{2}}m_{t_b}}\right)^{(J_1)}||A{-}1\, \alpha_1I_1T_1M_{T_1}}_{\rm SD}
\nonumber \\
   &\times
   \Braket{T_1M_{T_1}\textstyle{\frac{1}{2}}m_t|T_iM_{T_i}}\Braket{T_1'M_{T_1}'\textstyle{\frac{1}{2}}m_t'|T_fM_{T_f}}
   \nonumber\\ &+ \delta_{nn'}\delta_{\ell\ell'}
   \frac{\hat{s}\hat{s}'\hat{J_f}\hat{J_i}}{\hat{\kappa}}(-1)^{I_1'+l-{\textstyle\frac{1}{2}}+J_i+s+s'}
\begin{Bmatrix}
    s' & s & \kappa \\
    I_1 & I_1' & \textstyle{\frac{1}{2}} \\
\end{Bmatrix}
\begin{Bmatrix}
    s' & s & \kappa \\
    J_i & J_f & \ell \\
\end{Bmatrix}
\sum \Braket{j_a\textstyle{\frac{1}{2}}m_{t_a}||\hat{\mathcal{O}}^{(\kappa\tau)}_{\mu_\tau}||j_b\textstyle{\frac{1}{2}}m_{t_b}} \nonumber \\
   &\times _{\rm SD}\Braket{A{-}1\, \alpha_1'I_1'T_1'M_{T_1}'||\left[\hat{a}^\dagger_{j_a\textstyle{\frac{1}{2}}m_{t_a}}\tilde{a}_{j_b\textstyle{\frac{1}{2}}m_{t_b}}\right]^{(\kappa)}||A{-}1\, \alpha_1I_1T_1M_{T_1}}_{\rm SD} 
   \nonumber \\
   &\times
   \Braket{T_1M_{T_1}\textstyle{\frac{1}{2}}m_t|T_iM_{T_i}}
   \Braket{T_1'M_{T_1}'\textstyle{\frac{1}{2}}m_t|T_fM_{T_f}}
   \label{eq:rgm-kernel}
\end{align}
where $\Braket{\left[\left(\hat{a}^\dagger_{j_a\textstyle{\frac{1}{2}}m_{t_a}}\hat{a}^\dagger_{j\textstyle{\frac{1}{2}}m_t}\right)^{(J_2)}\left(\tilde{a}_{j'\textstyle{\frac{1}{2}}m_t'}\tilde{a}_{j_b\textstyle{\frac{1}{2}}m_{t_b}}\right)^{(J_4)}\right]^{(J_3)}}$, $\Braket{\left(\hat{a}^\dagger_{j_a\textstyle{\frac{1}{2}}m_{t_a}}\tilde{a}_{j'\textstyle{\frac{1}{2}}{m_t'}}\right)^{(J_2)}}$, $\Braket{\left(\hat{a}^\dagger_{j\textstyle{\frac{1}{2}}m_t}\tilde{a}_{j_b\textstyle{\frac{1}{2}}m_{t_b}}\right)^{(J_1)}}$, and $\Braket{\left(\hat{a}^\dagger_{j_a\textstyle{\frac{1}{2}}m_{t_a}}\tilde{a}_{j_b\textstyle{\frac{1}{2}}m_{t_b}}\right)^{(\kappa)}}$ are two- and one-body density matrices between the target (here $^{10}$Be) NCSM wave functions. The 12j symbol is the 12j(II) definition in Ref.~\cite{Varshalovich}.

The reduced matrix elements in Eqs.~\eqref{eq:coupling-kernel}, \eqref{eq:transpose}, \eqref{eq:rgm-kernel} are in the SD basis, thus they are not translationally invariant since they contain the spurious motion of the $(A{-}1)$-nucleon cluster center of mass. Before transforming from the SD-space to coordinate space, these matrix elements must be made translationally invariant. 
The procedure to remove this spurious motion in Ref.~\cite{Quaglioni:2009} is generalized here to be applicable to operators of non-zero order.
The SD cluster wavefunction is related to the invariant cluster wavefunction in the following way:
\begin{align}
   \Ket{\Phi_{\nu n}^{J^\pi T}}_{\rm SD} = &\sum_{n_r\ell_rNLJ_r}\hat{\ell}\hat{J_r}(-1)^{s+\ell_r+L+J}
\begin{Bmatrix}
    s & \ell_r & J_r \\
    L & J & \ell \\
\end{Bmatrix}
\braket{n_r\ell_rNL\ell|00n\ell\ell}_{\frac{a}{A-a}}
\left[\Ket{\Phi_{\nu_r n_r}^{J_r^{\pi_r} T}}\phi_{NL}(\vec{\xi}_0)\right]^{(J^\pi T)},
\label{eq:Jacobi}
\end{align}
where $\braket{n_r\ell_rNL\ell|00n\ell\ell}_{\frac{a}{A-a}}$ is a generalized HO bracket for two particles with mass ratio $d=\frac{a}{A-a}$~\cite{Trlifaj:1972}. In the present case of a single-nucleon projectile $a{=}1$. Using Eq.~\eqref{eq:Jacobi}, the translationally-invariant reduced matrix element can be extracted by inverting the following expression:
\begin{align}
   &_{\rm SD}\Braket{\Phi_{\nu'n'}^{J'^\pi T'M_T'}||\hat{\mathcal{A}}_{\nu'}\hat{\mathcal{O}}^{(\kappa\tau)}_{\mu_\tau}\hat{\mathcal{A}}_\nu||\Phi_{\nu n}^{J^\pi TM_T}}_{\rm SD} =  \sum_{NLn_r\ell_rJ_rn_r'\ell_r'J_r'}
   (-1)^{s+\ell_r+s'+\ell_r'+J'-J_r'+\kappa+L}
\hat{J}\hat{J'}\hat{\ell}\hat{\ell'}\hat{J_r}\hat{J_r'}
\begin{Bmatrix}
    \kappa & J_r & J_r' \\
    L & J' & J \\
\end{Bmatrix}
\nonumber\\
&\times
\begin{Bmatrix}
    s & \ell_r & J_r \\
    L & J & \ell \\
\end{Bmatrix}
\begin{Bmatrix}
    s' & \ell_r' & J_r' \\
    L & J' & \ell' \\
\end{Bmatrix}
\braket{n_r\ell_rNL\ell|00n\ell\ell}_{\frac{a}{A-a}}\braket{n_r'\ell_r'NL\ell'|00n'\ell'\ell'}_{\frac{a}{A-a}}
\nonumber\\
&\times\Braket{\Phi_{\nu_r'n_r'}^{J_r'^{\pi_r} T'M_T'}||\hat{\mathcal{A}}_{\nu'_r}\hat{\mathcal{O}}^{(\kappa\tau)}_{\mu_\tau}\hat{\mathcal{A}}_{\nu_r}||\Phi_{\nu_r n_r}^{J_r^{\pi_r} TM_T}}.
\label{eq:transform}
\end{align}
Equation~\eqref{eq:transform} is a generalization of Eq. (32) in Ref.~\cite{Quaglioni:2009}.

The removal of the spurious c.m. motion from the coupling matrix elements is still more straightforward. Following Refs.~~\cite{Simone:2013} and \cite{PhysRevC.70.054324}, we find 
\begin{align}
&_{\rm SD}\Braket{A\lambda J_fT_fM_{T_f}||\hat{\mathcal{O}}^{(\kappa\tau)}_{\mu_\tau}\hat{\mathcal{A}}_\nu||\Phi_{\nu n}^{J_iT_iM_{T_i}}}_{\rm SD} =
\langle n\ell 00\ell|00n\ell\ell\rangle_{\frac{a}{A-a}}
\Braket{A\lambda J_fT_fM_{T_f}||\hat{\mathcal{O}}^{(\kappa\tau)}_{\mu_\tau}\hat{\mathcal{A}}_\nu||\Phi_{\nu n}^{J_iT_iM_{T_i}}}
\label{eq:coupling-transform}
\end{align}
with a generalized HO bracket due to the c.m. motion, which value is simply given by
\begin{equation}\label{cm_ho_br}
\langle n\ell\ell00\ell|00n\ell\ell\rangle_{\frac{a}{A-a}} = (-1)^\ell \left(\frac{A-a}{A}\right)^{\frac{2n+\ell}{2}}
\; .
\end{equation}

After removing the spurious c.m. motion using either Eq.~\eqref{eq:transform} or ~\eqref{eq:coupling-transform}, the conversion of the matrix elements to coordinate space 
using Eq.~\eqref{eq:HO} is straight-forward with the exception of the first term in
Eq.~\eqref{eq:rgm-kernel}. The first term in Eq.~\eqref{eq:rgm-kernel} is completely contracted,
thus it contains the Kronecker delta $\delta_{nn'}$. The Kronecker delta would analytically convert
to a Dirac-delta $\delta(r-r')$ using Eq.~\eqref{eq:HO} assuming an infinitely large basis size. Of
course, in practice the HO basis is finite, so Eq.~\eqref{eq:HO} will not properly transform the
Kronecker delta from the first term in Eq.~\eqref{eq:rgm-kernel}. To account for this, we first
split the operator between the target and projectile:
\begin{equation}
   \hat{\mathcal{O}}_A = \hat{\mathcal{O}}_{(A-1)} + \hat{\mathcal{O}}_{1}.
   \label{eq:op_split}
\end{equation}
By splitting the operator, we can insert complete sets over the target and projectile NCSM states (note that in the rest of this section we omit isospin quantum numbers for simplicity):
\begin{align}
   &\Braket{\Phi_{\nu' r'}^{J_f}|\hat{A}_{\nu'}\hat{\mathcal{O}}^{(\kappa)}\hat{A}_{\nu}|\Phi_{\nu
   r}^{J_i}} = 
   \sum 
   \Braket{\Phi_{\nu' r'}^{J_f}|\hat{\mathcal{A}}_{\nu'}\hat{\mathcal{A}}_{\nu}|A{-}1\, \alpha_1I_1}
\nonumber \\
&\times 
   \Braket{A{-}1\, \alpha_1I_1|\hat{\mathcal{O}}^{(\kappa)}_{(A-1)}|\Phi_{\nu r}^{J_i}}
   +
   \sum 
   \Braket{\Phi_{\nu' r'}^{J_f}|\hat{\mathcal{A}}_{\nu'}\hat{\mathcal{A}}_{\nu}|\alpha_2I_2}
   \Braket{\alpha_2I_2|\hat{\mathcal{O}}^{(\kappa)}_{(1)}|\Phi_{\nu r}^{J_i}}.
\label{eq:op_split2}
\end{align}
After substituting Eq.~\eqref{eq:phi} and performing some angular momentum recoupling, the expression becomes:
\begin{align}
   &\Braket{\Phi_{\nu' r'}^{J_f}||\hat{A}_{\nu'}\hat{\mathcal{O}}^{(\kappa)}\hat{A}_{\nu}||\Phi_{\nu
   r}^{J_i}} = 
   \hat{J_i}\hat{J_f}\sum \hat{s}\hat{s}'
\begin{Bmatrix}
    \kappa & I_1' & I_1 \\
    I_2 & s & s' \\
\end{Bmatrix}
\begin{Bmatrix}
    \kappa & s' & s \\
    \ell & J_i & J_f \\
\end{Bmatrix}
(-1)^{J_i+\ell+s'-s-I_2-I_1'}
   \nonumber\\
   &\times
   \Braket{\Phi_{\nu' r'}^{J_f}|\hat{\mathcal{A}}_{\nu'}\hat{\mathcal{A}}_{\bar{\nu}}|\Phi_{\bar{\nu} r}^{J_f}} 
   \Braket{A{-}1\, \alpha_1'I_1'||\hat{\mathcal{O}}^{(\kappa}_{(A-1)}||A{-}1\, \alpha_1I_1}
\nonumber \\
&+ 
   \hat{J_i}\hat{J_f}\sum \hat{s}\hat{s}'
\begin{Bmatrix}
    \kappa & I_2' & I_2 \\
    I_1 & s & s' \\
\end{Bmatrix}
\begin{Bmatrix}
    \kappa & s' & s \\
    \ell & J_i & J_f \\
\end{Bmatrix}
(-1)^{J_i+\ell-I_2-I_1}
   \nonumber\\
   &\times
   \Braket{\Phi_{\nu' r'}^{J_f}|\hat{\mathcal{A}}_{\nu'}\hat{\mathcal{A}}_{\tilde{\nu}}|\Phi_{\tilde{\nu} r}^{J_f}} 
   \Braket{\alpha_2'I_2'||\hat{\mathcal{O}}^{(\kappa\tau)}_{(1)}||\alpha_2I_2} \; ,
\label{eq:delta1}
\end{align}
with the projectile states $\Ket{\alpha_2I_2T_2M_{T_2}}$ representing either a proton or a neutron with $\alpha_2{\equiv}1$, $I_2{=}1/2$. The cumulative quantum numbers $\bar{\nu}$ and $\tilde{\nu}$ represent $(A{-}1 \alpha_1' I_1' \alpha_2 I_2 s' \ell)$ and $(A{-}1 \alpha_1 I_1 \alpha_2' I_2' s' \ell)$, respectively.
The matrix elements of $\hat{\mathcal{A}}_{\nu'}\hat{\mathcal{A}}_{\bar{\nu}}$ are defined as the norm
kernels in Ref.~\cite{Navratil:2016}:

\begin{align}
   &\mathcal{N}_{\nu'\bar{\nu}}^{J_f}(r',r) = \Braket{\Phi_{\nu' r'}^{J_f}|\hat{\mathcal{A}}_{\nu'}\hat{\mathcal{A}}_{\bar{\nu}}|\Phi_{\bar{\nu} r}^{J_f}}
   \nonumber \\
   & = \delta_{\nu'\bar{\nu}}\frac{\delta(r'-r)}{r'r} +
   \sum_{nn'}R_{n'\ell'}(r')\left[N_{\nu'n'\bar{\nu} n}^{J_f}-\delta_{\nu'\bar{\nu}}\delta_{n'n}\right]
    R_{n\ell}(r), 
   \label{eq:norm}
\end{align}
where $N_{\nu'n'\bar{\nu} n}^{J_f} = \Braket{\Phi_{\nu'
n'}^{J_f}|\hat{\mathcal{A}}_{\nu'}\hat{\mathcal{A}}_{\bar{\nu}}|\Phi_{\bar{\nu}
n}^{J_f}}$ is the norm kernel in HO space. The coordinate-space norm kernel defined in Eq.~\eqref{eq:norm} is
written to explicitly treat the $\delta$-term in coordinate space and subtracting the completely contracted part
from the HO-space norm kernel. Using this expression, Eq.~\eqref{eq:delta1} becomes:
\begin{align}
   \Braket{\Phi_{\nu' r'}^{J_f}||\hat{\mathcal{A}}_{\nu'}\hat{\mathcal{O}}^{(\kappa)}\hat{\mathcal{A}}_{\nu}||\Phi_{\nu
   r}^{J_i}} &= 
   \hat{J_i}\hat{J_f}\sum \hat{s}\hat{s}'
\begin{Bmatrix}
    \kappa & I_1' & I_1 \\
    I_2 & s & s' \\
\end{Bmatrix}
\begin{Bmatrix}
    \kappa & s' & s \\
    \ell & J_i & J_f \\
\end{Bmatrix}
(-1)^{J_i+\ell+s'-s-I_2-I_1'}
   \nonumber\\
   &\times
   \delta_{\nu'\bar{\nu}}\frac{\delta(r'-r)}{r'r} 
   \Braket{A{-}1\, \alpha_1'I_1'||\hat{\mathcal{O}}^{(\kappa)}_{(A-1)}||A{-}1\, \alpha_1I_1}
\nonumber \\
&+ 
   \hat{J_i}\hat{J_f}\sum \hat{s}\hat{s}'
\begin{Bmatrix}
    \kappa & I_2' & I_2 \\
    I_1 & s & s' \\
\end{Bmatrix}
\begin{Bmatrix}
    \kappa & s' & s \\
    \ell & J_i & J_f \\
\end{Bmatrix}
(-1)^{J_i+\ell-I_2-I_1}
   \nonumber\\
   &\times
   \delta_{\nu'\tilde{\nu}}\frac{\delta(r'-r)}{r'r} 
   \Braket{\alpha_2'I_2'||\hat{\mathcal{O}}^{(\kappa)}_{(1)}||\alpha_2I_2}
   \nonumber\\
   &+ \sum_{nn'}R_{n'\ell'}(r')
   \Braket{\Phi_{\nu'n'}^{J_f}||\hat{\mathcal{A}}_{\nu'}\left[\hat{\mathcal{O}}^{(\kappa)}-1\right]\hat{\mathcal{A}}_{\nu}||\Phi_{\nu n}^{J_i}}
   R_{n\ell}(r).
\label{eq:delta2}
\end{align}
\end{widetext}
The HO matrix element in the last line of Eq.~\eqref{eq:delta2}, $\Braket{\Phi_{\nu'n'}^{J_f}||\hat{\mathcal{A}}_{\nu'}\left[\hat{\mathcal{O}}^{(\kappa)}-1\right]\hat{\mathcal{A}}_{\nu}||\Phi_{\nu n}^{J_i}}$, corresponds to Eq.~\eqref{eq:rgm-kernel} (after removing the c.m.~motion) with the completely-contracted term omitted.
\bibliographystyle{apsrev4-1}
\bibliography{beta}

\end{document}